\documentstyle[aps,tighten,preprint,epsfig,amssymb]{revtex}

\begin{document}

\draft

\preprint{nucl-th/0010064}


\title{
$\bbox{J/\psi}$ absorption by $\bbox{\pi}$ and $\bbox{\rho}$ mesons
in meson exchange model with anomalous parity interactions}


\author{
Yongseok Oh,%
\footnote{Email address: yoh@phya.yonsei.ac.kr}
Taesoo Song,%
\footnote{Email address: tssong@phya.yonsei.ac.kr}
and
Su Houng Lee\,%
\footnote{Email address: suhoung@phya.yonsei.ac.kr}}

\address{
Institute of Physics and Applied Physics and Department of Physics,
Yonsei University, Seoul 120-749, Korea}

\maketitle


\begin{abstract}
We reanalyze the dissociation process of the $J/\psi$ by $\pi$ and
$\rho$ mesons into $D + \bar{D}$, $D^* + \bar{D}$, $D + \bar{D}^*$,
and $D^* + \bar{D}^*$ within a meson exchange model.
In addition to the dissociation mechanisms considered in the literature,
we consider anomalous parity interactions, whose couplings are constrained
by heavy quark spin symmetry and phenomenology. 
This opens new dissociation channels and adds new diagrams in the 
previously considered processes.  
Compared to the previous results, we find that these new additions have
only a minor effect on the $\rho + J/\psi$ total inelastic cross section,
but reduce the one for $\pi + J/\psi$ by about 50~\% near the threshold.
\end{abstract}



\section{Introduction}

Suppression of $J/\psi$ production is considered to be one of the
promising signals for detecting the formation of the quark-gluon
plasma (QGP) in relativistic heavy ion collisions (RHIC) \cite{MS86,Satz00}.  
Indeed, recent reports by the NA50 Collaboration \cite{NA50-96,NA50-00}
show an anomalous suppression of $J/\psi$ production in Pb$+$Pb
collisions at CERN.  
Model calculations, which assume the formation of a deconfined state of
quarks and gluons \cite{BO96a,Wong98a} at the initial stages of the
collision, seem to be successful in explaining the anomalous suppression.  
However, before coming to a definite conclusion on the existence of the 
QGP state in such collisions, it is essential to investigate whether the
observed $J/\psi$ suppression can be explained by a more conventional
approach that does not assume the existence of QGP state.
In such models, the suppression comes from the absorption of $J/\psi$
by the co-moving hadrons \cite{comove,CKo97,CB97,AC98}.
Since such co-mover models strongly depend on the cross sections of
$J/\psi$ absorption by hadrons, it is essential to have a better knowledge
of the magnitude and energy dependence of the cross sections.  
This is not an easy task, but is indispensable for true identification of 
QGP in RHIC.

The absorption of $J/\psi$ by light hadrons has been investigated using
various methods \cite{KSa94,KSSZ96,MBQ95,WSB00b,MMu98,Hagl00,LK00,HG00b}.
The estimated values for the cross sections show a strong dependence on
the hadronic models and assumptions on the absorption mechanisms employed
in the calculation.
The first set of models uses the quark degrees of freedom.
Using perturbative QCD in the heavy charm quark mass limit \cite{Peskin79}, 
Kharzeev, Satz, and their collaborators \cite{KSa94,KSSZ96} found very
small absorption cross sections at the order of $\mu$b.
In Ref. \cite{MBQ95}, Martins, Blaschke, and Quack investigated the
charmonium dissociation processes, $\pi + J/\psi \to D^* + \bar{D}, D +
\bar{D}^*$, and $D^* + \bar{D}^*$ using quark exchange model and found
a strong enhancement near the threshold with an exponential falloff at
higher energies.
The peak was found at $\sqrt{s} \simeq 4$ GeV with a value of about $7$ mb.
This calculation was recently improved by Wong, Swanson, and Barnes
\cite{WSB00b} by using a more successful quark-interchange model and
hadron wave functions.
The cross section for $\pi + J/\psi$ was found to be relatively small with
a maximum of about $1$ mb, while the pion induced $\psi'$ dissociation
process had a much larger cross section with lower threshold for the
initial kinetic energy.

Another approach to $J/\psi$ absorption is to use effective meson
Lagrangian \cite{MMu98,Hagl00,LK00,HG00b}.
In this approach, the authors considered $t$-channel meson exchanges to
estimate the $J/\psi$ absorption cross sections.
In Ref. \cite{MMu98}, Matinyan and M{\"u}ller calculated $D$ meson
exchange diagrams and found small cross sections around $0.3$ mb for
$\pi + J/\psi$ and $\rho + J/\psi$.
The obtained cross sections were found to increase with energy.
This calculation was improved by Haglin \cite{Hagl00}, who included $D^*$
meson exchanges and four-point couplings in the $J/\psi$ absorption
mechanisms.
Similar calculations were recently performed by Lin and Ko \cite{LK00}.
The new results show a rather large cross section for $\pi + J/\psi \to
D^* + \bar{D}$ and $D + \bar{D}^*$, i.e., about $20$ mb at $\sqrt{s} = 4$
GeV, which at higher energies saturates to about $30$ mb.%
\footnote{The typo errors in Ref. \cite{Hagl00} and different
conventions between Refs. \cite{Hagl00,LK00} are cleared recently by
Ref. \cite{HG00b}.}
However, for a realistic structure of the mesons, form factors have to be
taken into account.   
Inclusion of form factors generally reduces the cross sections.
With suitable form factors and cutoff parameters, which however cannot
be justified {\em a priori\/}, Lin and Ko \cite{LK00} concluded that the
saturated absorption cross sections for $\pi + J/\psi$ and $\rho + J/\psi$
are around $7$ mb and $3$ mb, respectively.

All of the meson exchange model calculations discussed above essentially
use the same effective meson Lagrangian; namely the minimal SU(4)
Yang-Mills Lagrangian.  
Two immediate questions can arise.
The first is the use of SU(4) symmetry in constructing the effective
Lagrangian.
Obviously, charm quark is much heavier than the other three light quarks
and the SU(4) symmetry is nowhere near being true in QCD.
Nevertheless, the starting point of using the SU(4) Lagrangian is to
categorize the possible interaction vertices among the meson multiplets
and estimate as many of their respective couplings by phenomenology as
possible.
Doing so, not all of the couplings can be fixed.  
However, it turned out that for couplings that can be checked, one finds 
that the SU(4) symmetry relations for coupling constants are not totally
meaningless \cite{LK00,HG00b}.
Such checks provide the order of uncertainty for the calculated absorption
cross sections within the effective Lagrangian method, and makes the
calculation meaningful.  
The second question is whether all possible interactions are considered.  
It is to this question that we want to address in this paper.  
The previous studies did not fully take into account all possible set of  
interaction Lagrangian involving $D$ and $D^*$.  
In fact, the anomalous parity interactions which are connected to the
gauged Wess-Zumino action are all missing.   
Furthermore, the anomalous term such as $D^* D^* \pi$ is essential and
required by heavy quark spin symmetry and its coupling is strongly
constrained to the $D^* D \pi$ coupling \cite{Wise92,YCCL92}.
As we shall see, inclusion of $D^* D^* \pi$ coupling opens new absorption
channels and mechanisms which were not considered in the literature.
Therefore, to fully estimate the $J/\psi$ absorption cross section within
the hadronic model, it is necessary and important to see the effect
of the anomalous interactions to such processes, which we will investigate
in this work.

This paper is organized as follows. In Sec. II, we introduce our
effective Lagrangian. The interaction Lagrangian needed for $J/\psi$
absorption processes by $\pi$ and $\rho$ mesons are obtained and the
cross sections are evaluated.
Then in Sec. III, we determine the coupling constants and give the
numerical results for the cross sections.
Section IV contains summary and discussions.
Some details on deriving our effective Lagrangian are given in
Appendix.

\section{Effective Lagrangian for $J/\psi$ absorption}

In the heavy quark mass limit ($m_Q \rightarrow \infty$), effective
Lagrangian for light ($\bar{q} q$) and heavy ($\bar{q} Q$ or
$\bar{Q} q$) mesons can be constructed to preserve chiral and
heavy quark symmetries \cite{Wise92,YCCL92,VS87-88,IW89-90,Chan97}.  
When the heavy quark flavor is the charm, finite mass corrections are
expected to be important and should be introduced in the effective
Lagrangian in a systematic way.  
The couplings of the finite mass corrections can be fixed by
``velocity reparameterization invariance'' \cite{VRIs} or by phenomenology.
This ``top-down'' approach seems to be most plausible way to
construct the effective Lagrangian.
However, it is not yet clear how to implement heavy quark symmetry
to the interactions of quarkonia with light hadrons as the quarkonium
states ($\bar{Q} Q$) such as the $J/\psi$ contain two heavy constituents.

Another approach which was used in previous investigations
\cite{MMu98,Hagl00,LK00,HG00b} is the ``bottom-up'' approach.
In this approach, one starts with the SU(4) symmetric chiral Lagrangian.
The heavy charm quark mass  is then assumed to be taken into account
through appropriate symmetry breaking terms \cite{MOPR95}.  
However, since the SU(4) symmetry is badly broken, finding all the
appropriate symmetry breaking terms is very hard and beyond the scope
of this work.
In this paper, therefore, we use the ``bottom-up'' approach
phenomenologically to categorize all the possible interaction vertices
among the meson multiplets.  
The couplings are then determined by experimental data and by
implementing heavy quark spin symmetry.
Doing so, most of the three point couplings can be determined.
For the four point couplings and some three point couplings, where
such determination is not possible, we will partly rely on the SU(4)
relations as well as other model predictions.

As in Refs. \cite{Hagl00,LK00,HG00b}, we start with the SU(4) Lagrangian
for the pseudoscalar mesons and introduce the vector and axial-vector
mesons by the massive Yang-Mills approach \cite{Mei88}.
The axial-vector fields can be gauged-away as in Ref. \cite{KS85}.
The procedure to obtain the effective Lagrangian is well explained,
e.g., in Refs. \cite{LK00,HG00b} and is summarized in Appendix.
The effective Lagrangian for $\pi$, $\rho$, $D$, and $D^*$ then reads
\begin{mathletters} \label{Lag1}
\begin{eqnarray}
{\mathcal{L}}_{D^* D \pi} &=& i g_{D^*D\pi}^{} \left( D_\mu^* \partial^\mu
\pi \bar{D} - D \partial^\mu \pi \bar{D}^*_\mu \right),
\\
\mathcal{L}_{\psi DD} &=& i g_{\psi DD}^{} \psi_\mu \left( \partial^\mu
D
\bar{D} - D \partial^\mu \bar{D} \right),
\\
{\mathcal{L}}_{\psi D^* D^*} &=& -i g_{\psi D^* D^*}^{} \Bigl\{
\psi^\mu \left( \partial_\mu D^{*\nu} \bar{D}_\nu^* - D^{*\nu}
\partial_\mu \bar{D}_\nu^* \right) 
+ \left( \partial_\mu \psi_\nu D^{*\nu} - \psi_\nu \partial_\mu D^{*\nu}
\right) \bar{D}^{*\mu}
\nonumber \\ && \mbox{} \qquad\quad
+ D^{*\mu} \left( \psi^\nu \partial_\mu \bar{D}^*_\nu - \partial_\mu
\psi_\nu \bar{D}^{*\nu} \right)
\Bigr\},
\\
{\mathcal{L}}_{\psi D^* D\pi} &=& g_{\psi D^* D\pi}^{} \psi^\mu \left( D
\pi \bar{D}^*_\mu + D^*_\mu \pi \bar{D} \right),
\\
{\mathcal{L}}_{DD\rho} &=& i g_{DD\rho}^{} \left( D \rho^\mu \partial_\mu
\bar{D} - \partial_\mu D \rho^\mu \bar{D} \right),
\\
{\mathcal{L}}_{D^*D^*\rho} &=& i g_{D^*D^*\rho}^{}  \Bigl\{
\partial_\mu D^*_\nu \rho^\mu \bar{D}^{*\nu} - D_\nu^* \rho_\mu
\partial^\mu \bar{D}^{*\nu}
+ \left( D^{*\nu} \partial_\mu \rho_\nu - \partial_\mu D_\nu^* \rho^\nu
\right) \bar{D}^{*\mu}
\nonumber \\ && \mbox{} \qquad\quad
+ D^{*\mu} \left( \rho^\nu \partial_\mu \bar{D}^*_\nu - \partial_\mu
\rho_\nu \bar{D}^{*\nu} \right) \Bigr\},
\\
{\mathcal{L}}_{\psi DD\rho} &=& -g_{\psi DD\rho}^{} \psi^\mu D \rho_\mu
\bar{D},
\\
{\mathcal{L}}_{\psi D^* D^* \rho} &=& g_{\psi D^* D^* \rho}^{} \psi_\mu
\left( 2 D^{*\nu} \rho^\mu \bar{D}^*_\nu - D^{*\nu} \rho_\nu
\bar{D}^{*\mu} - D^{*\mu} \rho^\nu \bar{D}^*_\nu \right),
\end{eqnarray}
\end{mathletters}
where $\pi = \bbox{\tau} \cdot \bbox{\pi}$, $\rho = \bbox{\tau} \cdot
\bbox{\rho}$, and we define the charm meson iso-doublets as
\begin{eqnarray}
&& \bar{D}^T = \left( \bar{D}^0 ,  D^- \right),
\qquad D = \left( D^0 , D^+ \right),
\nonumber \\
&& \bar{D}^{*T} = \left( \bar{D}^{*0} ,  D^{*-} \right),
\qquad D^* = \left( D^{*0} , D^{*+} \right).
\end{eqnarray}

In addition to the normal terms given above which were used by Refs.
\cite{Hagl00,LK00,HG00b}, there are anomalous parity terms.
The anomalous parity interactions with vector fields can be described in
terms of the gauged Wess-Zumino action \cite{KRS84} and are summarized
in Appendix.  
By the same method used to obtain the normal terms, we get
\begin{mathletters} \label{Lag2}
\begin{eqnarray}
{\mathcal{L}}_{D^* D^* \pi} &=& -g_{D^* D^* \pi}^{}
\varepsilon^{\mu\nu\alpha\beta} \partial_\mu D^*_\nu \pi \partial_\alpha
\bar{D}^*_\beta,
\\
{\mathcal{L}}_{\psi D^* D} &=& -g_{\psi D^* D}^{}
\varepsilon^{\mu\nu\alpha\beta} \partial_\mu \psi_\nu \left(
\partial_\alpha D^*_\beta \bar{D} + D \partial_\alpha \bar{D}^*_\beta
\right),
\\
{\mathcal{L}}_{\psi DD \pi} &=& -i g_{\psi DD \pi}^{}
\varepsilon^{\mu\nu\alpha\beta} \psi_\mu \partial_\nu D \partial_\alpha
\pi \partial_\beta \bar{D},
\\
{\mathcal{L}}_{\psi D^* D^* \pi} &=& -i g_{\psi D^* D^* \pi}^{}
\varepsilon^{\mu\nu\alpha\beta} \psi_\mu D^*_\nu \partial_\alpha \pi
\bar{D}^*_\beta
- i h_{\psi D^* D^* \pi}^{} \varepsilon^{\mu\nu\alpha\beta} \partial_\mu
  \psi_\nu D_\alpha^* \pi \bar{D}_\beta^*,
\\
{\mathcal{L}}_{D^* D \rho} &=& -g_{D^* D \rho}^{}
\varepsilon^{\mu\nu\alpha\beta} \left( D \partial_\mu \rho_\nu
\partial_\alpha \bar{D}^*_\beta + \partial_\mu D^*_\nu \partial_\alpha
\rho_\beta \bar{D} \right),
\\
{\mathcal{L}}_{\psi D^* D \rho} &=& i g_{\psi D^* D \rho}^{}
\varepsilon^{\mu\nu\alpha\beta} \psi_\mu \left( \partial_\nu D
\rho_\alpha \bar{D}^*_\beta + D^*_\nu \rho_\alpha \partial_\beta \bar{D}
\right)
\nonumber \\ && \mbox{}
- i h_{\psi D^* D \rho}^{} \varepsilon^{\mu\nu\alpha\beta} \psi_\mu
  \left( D \rho_\nu \partial_\alpha \bar{D}_\beta^* - \partial_\nu
D^*_\alpha \rho_\beta \bar{D} \right),
\end{eqnarray}
\end{mathletters}
with $\varepsilon_{0123} = +1$.
In the heavy mass limit, due to the heavy quark spin symmetry, the
pseudoscalars and vectors become degenerate and have to be treated on
the same footing.  
As a consequence, the anomalous $D^* D^* \pi$ interaction term is required
and can be related to the  $D^* D \pi$ interaction term by the heavy
quark spin symmetry \cite{Wise92,YCCL92}.

\begin{figure}[h]
\centering
\epsfig{file=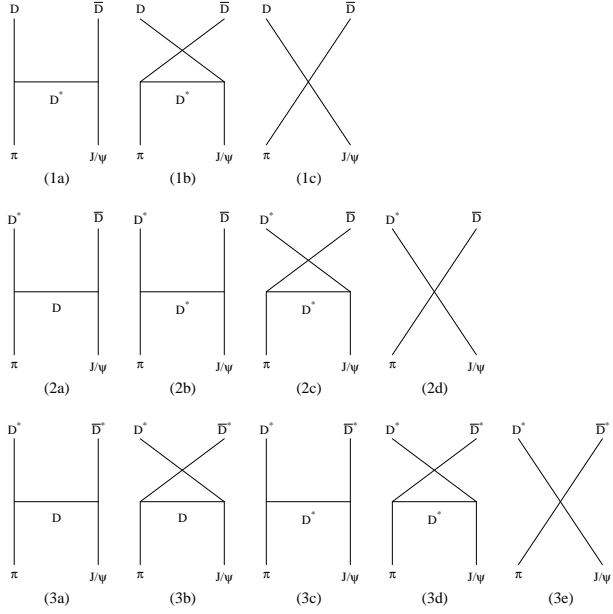, width=0.5\hsize}
\caption{Feynman diagrams for $J/\psi$ absorption processes by pion:
(1) $J/\psi + \pi \to D + \bar{D}$,
(2) $J/\psi + \pi \to D^* + \bar{D}$, and
(3) $J/\psi + \pi \to D^* + \bar{D}^*$.
The process $J/\psi + \pi \to D + \bar{D}^*$ has the same cross section
as (2).}
\label{fig:pions}
\end{figure}

\begin{figure}[h]
\centering
\epsfig{file=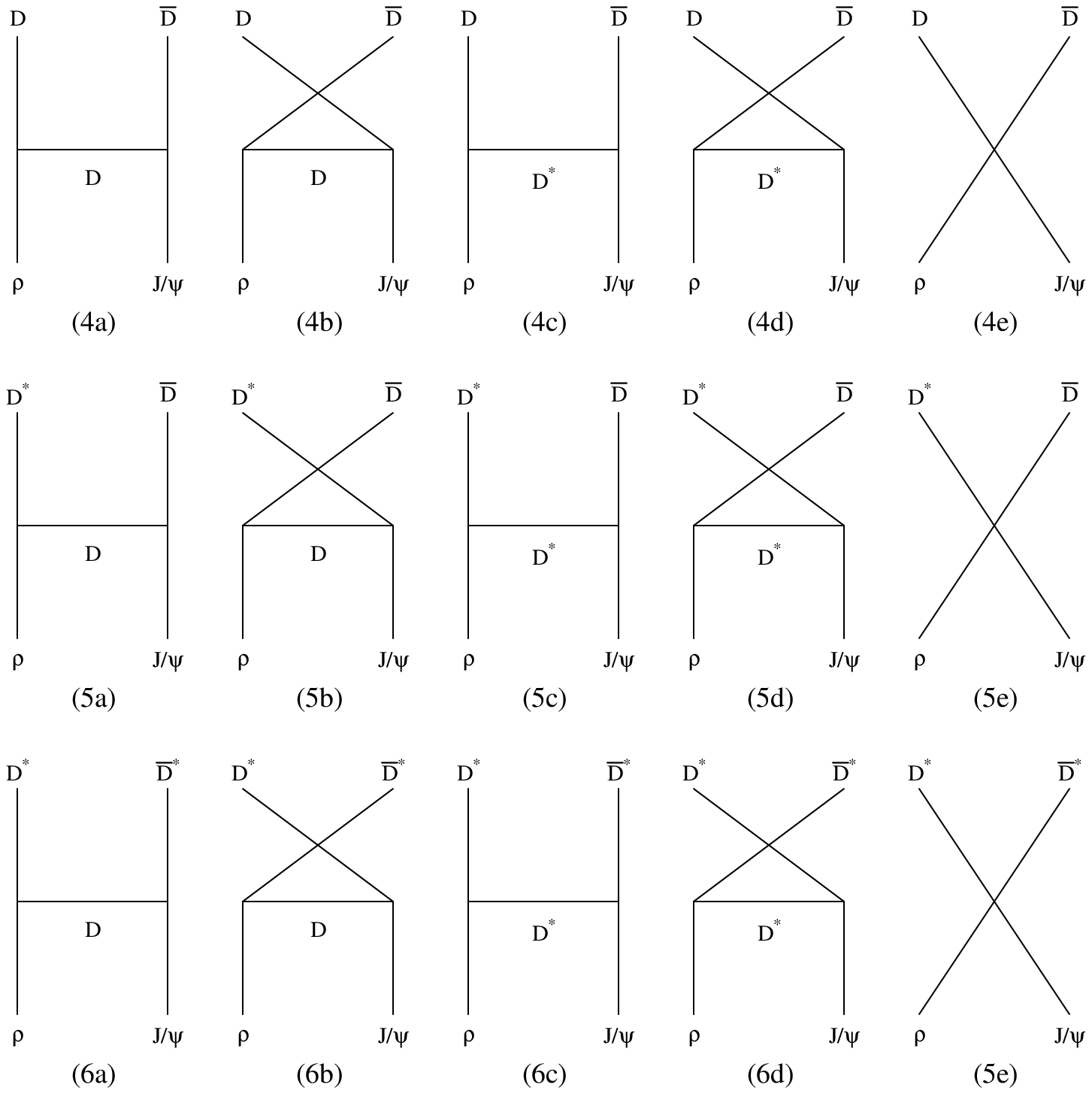, width=0.5\hsize}
\caption{Feynman diagrams for $J/\psi$ absorption processes by $\rho$:
(4) $J/\psi + \rho \to D + \bar{D}$,
(5) $J/\psi + \rho \to D^* + \bar{D}$, and
(6) $J/\psi + \rho \to D^* + \bar{D}^*$.
The process $J/\psi + \rho \to D + \bar{D}^*$ has the same cross section
as (5).}
\label{fig:rhos}
\end{figure}

Let us first evaluate the absorption processes of the $J/\psi$ by $\pi$
and $\rho$ mesons within the effective Lagrangian obtained above.
The values of the coupling constants will be discussed later.   
In this study, we assume that the $J/\psi$ is a pure $\bar{c}c$ state
and consider the OZI-preserving processes only at the tree level.
The diagrams we are calculating are shown in Figs. \ref{fig:pions} and
\ref{fig:rhos}.
Compared to the previous studies \cite{Hagl00,LK00}, we find that
the anomalous parity interactions open new absorption channels, namely
$\pi + J/\psi \to D + \bar{D}$ [Diagram (1)],
$\pi + J/\psi \to D^* + \bar{D}^*$ [Diagram (3)], and
$\rho + J/\psi \to D^* + \bar{D}$ [Diagram (5)].
These processes are not allowed by the normal terms and were not
considered before in the effective Lagrangian approach.
In the quark exchange model \cite{MBQ95,WSB00b}, however, the processes
$\pi + J/\psi \to D^* + \bar{D}^*$ and $\rho + J/\psi \to D^* + \bar{D}$
were calculated.
There, $\pi + J/\psi \to D + \bar{D}$ was not considered since the
absorption amplitude vanishes unless one considers the spin-orbit force
\cite{WSB00b,BBIK00}.
A recent study using the Dyson-Schwinger formalism shows that the 
cross section for $\pi + J/\psi \to D + \bar{D}$ is small \cite{BBIK00}.
The anomalous terms also allow new absorption mechanisms to the previously
considered absorption channels.
These are Diagrams (2b), (4c), (4d), (6a), and (6b) of
Figs. \ref{fig:pions} and \ref{fig:rhos}.
As we shall see, the role of those diagrams is not negligible especially
for the $J/\psi$ absorption by the $\pi$'s.

We first calculate the absorption amplitude of $\pi + J/\psi$.
We define the four-momenta of the $J/\psi$ and $\pi$ by $p_1$ and $p_2$,
respectively.
We also denote the four-momenta of the final particles by $p_3$ and
$p_4$, which then defines $s=(p_1+p_2)^2$ and $t=(p_1-p_3)^2$.
The polarization vector of the vector meson with momentum $p_i$ is
represented by $\varepsilon_i$.
$M_D$ and $M_{D^*}$ represent the $D$ and $D^*$ meson masses, respectively.  
Then the differential cross section for this process is
\begin{equation}
\frac{d\sigma}{dt} = \frac{1}{96\pi s {\bf p}_1^2}
\sum_{\rm spin} | {\mathcal{M}} |^2,
\label{dsigdt}
\end{equation}
where ${\bf p}_1$ is the three-momentum of $p_1$ in the center of mass frame.
Note that we have multiplied factor $2$ to get Eq. (\ref{dsigdt}).  
This comes from summing over the possible isospin quantum numbers of the 
final state \cite{LK00}.
In computing Diagram (1) in Fig. \ref{fig:pions}, the four-momenta of
the $D$ and $\bar{D}$ are represented by $p_3$ and $p_4$, respectively.  
The absorption amplitude for this diagram can be written as
\begin{equation}
{\mathcal{M}} = \sum_i {\mathcal{M}}_\mu^{(1i)} \varepsilon_1^\mu,
\end{equation}
where 
\begin{eqnarray}
{\mathcal{M}}^{(1a)}_\mu &=& - \frac{g_{D^* D \pi}^{} g_{\psi D^*
D}^{}}{(p_2-p_3)^2 - M_{D^*}^2} \varepsilon_{\mu\beta\lambda\rho}
\left\{ g^{\alpha\beta} - \frac{(p_2-p_3)^\alpha
(p_2-p_3)^\beta}{M_{D^*}^2} \right\} p_{2\alpha} p_1^\lambda p_4^\rho,
\nonumber \\
{\mathcal{M}}^{(1b)}_\mu &=& - \frac{g_{D^* D \pi}^{} g_{\psi D^*
D}^{}}{(p_1-p_3)^2 - M_{D^*}^2} \varepsilon_{\mu\beta\lambda\rho}
\left\{ g^{\alpha\beta} - \frac{(p_1-p_3)^\alpha
(p_1-p_3)^\beta}{M_{D^*}^2} \right\}  p_{2\alpha} p_1^\lambda p_3^\rho,
\nonumber \\
{\mathcal{M}}^{(1c)}_\mu &=& g_{\psi DD \pi}
\varepsilon_{\mu\beta\lambda\rho} p_2^\beta p_3^\lambda p_4^\rho.
\end{eqnarray}

In Diagram (2) of Fig. \ref{fig:pions}, the four-momenta of the $D^*$ and
$\bar{D}$ are denoted by $p_3$ and $p_4$, respectively, and we have 
\begin{equation}
{\mathcal{M}} = \varepsilon_3^{*\nu} \sum_i {\mathcal{M}}_{\mu\nu}^{(2i)}
\varepsilon_1^\mu,
\end{equation}
where
\begin{eqnarray}
{\mathcal{M}}^{(2a)}_{\mu\nu} &=& \frac{2 g_{D^* D\pi}^{} g_{\psi
DD}^{}}{(p_2-p_3)^2 - M_D^2} p_{2\nu} p_{4\mu},
\nonumber \\
{\mathcal{M}}^{(2b)}_{\mu\nu} &=& - \frac{ g_{D^* D^* \pi}^{} g_{\psi D^*
D}^{}}{(p_2-p_3)^2 - M_{D^*}^2} \varepsilon_{\mu\gamma\delta\beta}
\varepsilon_{\nu\lambda\rho\alpha} \left\{ g^{\alpha\beta} -
\frac{(p_2-p_3)^\alpha (p_2-p_3)^\beta}{M_{D^*}^2} \right\}
p_1^\gamma p_2^\rho p_3^\lambda p_4^\delta,
\nonumber \\
{\mathcal{M}}^{(2c)}_{\mu\nu} &=& -\frac12 \frac{g_{D^* D \pi}^{}
g_{\psi D^* D^*}^{}}{(p_1-p_3)^2 - M_{D^*}^2} \left\{ g^{\alpha\beta} -
\frac{(p_1-p_3)^\alpha (p_1-p_3)^\beta}{M_{D^*}^2} \right\}
\nonumber \\ && \mbox{} \times
\left( p_{2\alpha} + p_{4\alpha} \right)
\left[ 2p_{3\mu} g_{\nu\beta} - (p_1 + p_3)_\beta
g_{\mu\nu} + 2p_{1\nu} g_{\mu\beta} \right],
\nonumber \\
{\mathcal{M}}^{(2d)}_{\mu\nu} &=& - g_{\psi D^* D \pi}^{} g_{\mu\nu}.
\end{eqnarray}

For Diagram (3) of Fig. \ref{fig:pions}, we write the four-momenta of
the $D^*$ and $\bar{D}^*$ by $p_3$ and $p_4$, respectively, and we have 
\begin{equation}
{\mathcal{M}} = \varepsilon_4^{*\lambda} \varepsilon_3^{*\nu}
\sum_i {\mathcal{M}}_{\mu\nu\lambda}^{(3i)} \varepsilon_1^\mu,
\end{equation}
where
\begin{eqnarray}
{\mathcal{M}}^{(3a)}_{\mu\nu\lambda} &=& \frac{g_{D^* D \pi}^{} g_{\psi
D^*
D}^{}}{(p_2 - p_3)^2 - M_D^2} \varepsilon_{\mu\lambda\gamma\delta}\,
p_1^\gamma p_{2\nu} p_4^\delta,
\nonumber \\
{\mathcal{M}}^{(3b)}_{\mu\nu\lambda} &=& - \frac{g_{D^* D\pi}^{} g_{\psi
D^* D}^{}}{(p_1-p_3)^2 - M_D^2} \varepsilon_{\mu\nu\gamma\delta}\,
p_1^\gamma p_{2\lambda} p_3^\delta,
\nonumber \\
{\mathcal{M}}^{(3c)}_{\mu\nu\lambda} &=& - \frac{g_{D^* D^* \pi}^{}
g_{\psi
D^* D^*}^{}}{(p_2-p_3)^2 - M_{D^*}^2} \left\{ g^{\alpha\beta} -
\frac{(p_2-p_3)^\alpha (p_2-p_3)^\beta}{M_{D^*}^2} \right\}
\nonumber \\ && \mbox{} \times
\varepsilon_{\nu\gamma\delta\alpha} p_3^\gamma p_2^\delta \left[
2p_{4\mu} g_{\beta\lambda} - (p_1 + p_4)_\beta g_{\mu\lambda}
+ 2p_{1\lambda} g_{\mu\beta} \right],
\nonumber \\
{\mathcal{M}}^{(3d)}_{\mu\nu\lambda} &=& - \frac{g_{D^* D^* \pi}^{}
g_{\psi
D^* D^*}^{}}{(p_1-p_3)^2 - M_{D^*}^2} \left\{ g^{\alpha\beta} -
\frac{(p_1-p_3)^\alpha (p_1-p_3)^\beta}{M_{D^*}^2} \right\}
\nonumber \\ && \mbox{} \times
\varepsilon_{\lambda\gamma\delta\alpha} p_2^\gamma p_4^\delta \left[
2p_{3\mu} g_{\beta\nu} - (p_1+p_3)_\beta g_{\mu\nu} +
2p_{1\nu} g_{\mu\beta} \right],
\nonumber \\
{\mathcal{M}}^{(3e)}_{\mu\nu\lambda} &=& - g_{\psi D^* D^* \pi}^{}
\varepsilon_{\mu\nu\lambda\rho} p_2^\rho - h_{\psi D^* D^* \pi}^{}
\varepsilon_{\mu\nu\lambda\rho} p_1^\rho.
\end{eqnarray}

We now calculate the $J/\psi$ absorption by the $\rho$ meson as shown in Fig.
\ref{fig:rhos}.
We first define the four-momenta of the $J/\psi$ and $\rho$ as $p_1$ and
$p_2$ respectively.
Then the differential cross section reads
\begin{equation}
\frac{d\sigma}{dt} = \frac{1}{288\pi s {\bf p}_1^2}
\sum_{\rm spin} | {\mathcal{M}} |^2.
\end{equation}
For Diagram (4) of Fig. \ref{fig:rhos}, we define $p_3$ and $p_4$ as
the four-momenta of $D$ and $\bar{D}$ respectively.   Then,
\begin{equation}
{\mathcal{M}} = \sum_i {\mathcal{M}}^{(4i)}_{\mu\nu}
\varepsilon_1^\mu \varepsilon_2^\nu,
\end{equation}
where
\begin{eqnarray}
{\mathcal{M}}^{(4a)}_{\mu\nu} &=& \frac{4g_{DD\rho}^{} g_{\psi DD}^{}}{
(p_2-p_3)^2 - M_D^2} p_{3\nu} p_{4\mu},
\nonumber \\
{\mathcal{M}}^{(4b)}_{\mu\nu} &=& \frac{4g_{DD\rho}^{} g_{\psi DD}^{}}{
(p_1-p_3)^2 - M_D^2} p_{4\nu} p_{3\mu},
\nonumber \\
{\mathcal{M}}^{(4c)}_{\mu\nu} &=& - \frac{g_{D^* D^* \rho}^{} g_{\psi D^*
D}^{}}{(p_2-p_3)^2 - M_{D^*}^2} \left\{ g^{\alpha\beta} -
\frac{(p_2-p_3)^\alpha (p_2-p_3)^\beta}{M_{D^*}^2} \right\}
\varepsilon_{\nu\gamma\delta\alpha} \, \varepsilon_{\mu\lambda\rho\beta}
\, p_1^\lambda p_2^\gamma p_3^\delta p_4^\rho,
\nonumber \\
{\mathcal{M}}^{(4d)}_{\mu\nu} &=& \frac{g_{D^* D\rho}^{} g_{\psi D^*
D}^{}}{(p_1-p_3)^2 - M_{D^*}^2} \left\{ g^{\alpha\beta} -
\frac{(p_1-p_3)^\alpha (p_2-p_3)^\beta}{M_{D^*}^2} \right\}
\varepsilon_{\nu\gamma\delta\alpha} \, \varepsilon_{\mu\lambda\rho\beta}
\, p_1^\lambda p_2^\delta p_3^\rho p_4^\gamma,
\nonumber \\
{\mathcal{M}}^{(4e)}_{\mu\nu} &=& g_{\psi DD\rho}^{} g_{\mu\nu}.
\end{eqnarray}

For Diagram (5) of Fig. \ref{fig:rhos}, $p_3$ and $p_4$ are defined to
be the four-momenta of the $D^*$ and $\bar{D}$ respectively.
We have 
\begin{equation}
{\mathcal{M}} = \varepsilon_3^{*\lambda} \sum_i
{\mathcal{M}}^{(5i)}_{\mu\nu\lambda}
\varepsilon_1^\mu \varepsilon_2^\nu,
\end{equation}
where
\begin{eqnarray}
{\mathcal{M}}^{(5a)}_{\mu\nu\lambda} &=& - \frac{2 g_{D^* D \rho}^{}
g_{\psi DD}^{}}{(p_2-p_3)^2 - M_D^2}
\varepsilon_{\nu\lambda\gamma\delta}
\, p_2^\delta p_3^\gamma p_{4\mu},
\nonumber \\
{\mathcal{M}}^{(5b)}_{\mu\nu\lambda} &=& \frac{2 g_{D D\rho}^{} g_{\psi
D^* D}^{}}{(p_1-p_3)^2 - M_D^2} \varepsilon_{\lambda\mu\gamma\delta}
p_1^\gamma p_3^\delta p_{4\nu},
\nonumber \\
{\mathcal{M}}^{(5c)}_{\mu\nu\lambda} &=& - \frac{ g_{D^* D^* \rho}^{}
g_{\psi D^* D}^{}}{(p_2-p_3)^2 - M_{D^*}^2} \left\{ g^{\alpha\beta} -
\frac{(p_2-p_3)^\alpha (p_2-p_3)^\beta}{M_{D^*}^2} \right\}
\nonumber \\ && \mbox{} \times
\varepsilon_{\mu\gamma\delta\beta} p_1^\gamma p_4^\delta \left[ 2
p_{3\nu} g_{\lambda\alpha} - (p_2 + p_3)_\alpha g_{\nu\lambda} + 2
p_{2\lambda} g_{\nu\alpha} \right],
\nonumber \\
{\mathcal{M}}^{(5d)}_{\mu\nu\lambda} &=& - \frac{g_{D^* D\rho}^{} g_{\psi
D^* D^*}^{}}{(p_1-p_3)^2 - M_{D^*}^2} \left\{ g^{\alpha\beta} -
\frac{(p_1-p_3)^\alpha (p_1-p_3)^\beta}{M_{D^*}^2} \right\}
\nonumber \\ && \mbox{} \times
\varepsilon_{\nu\gamma\delta\alpha} p_2^\delta p_4^\gamma \left[ 2
p_{3\mu} g_{\beta\lambda} - (p_1 + p_3)_\beta g_{\mu\lambda} + 2
p_{1\lambda} g_{\mu\beta} \right],
\nonumber \\
{\mathcal{M}}^{(5e)}_{\mu\nu\lambda} &=& - g_{\psi D^* D \rho}^{}
\varepsilon_{\mu\nu\lambda\delta} p_4^\delta + h_{\psi D^* D \rho}^{}
\varepsilon_{\mu\nu\lambda\delta} p_3^\delta.
\end{eqnarray}

Finally for Diagram (6) of Fig. \ref{fig:rhos}, we denote the
four-momenta of the $D^*$ and $\bar{D}^*$ by $p_3$ and $p_4$,
respectively, then we have
\begin{equation}
{\mathcal{M}} = \varepsilon_4^{*\rho} \varepsilon_3^{*\lambda}
\sum_i {\mathcal{M}}^{(6i)}_{\mu\nu\lambda\rho}
\varepsilon_1^\mu \varepsilon_2^\nu,
\end{equation}
and
\begin{eqnarray}
{\mathcal{M}}^{(6a)}_{\mu\nu\lambda\rho} &=& - \frac{g_{D^* D\rho}^{}
g_{\psi D^* D}^{}}{(p_2-p_3)^2 - M_D^2} \,
\varepsilon_{\nu\lambda\gamma\delta} \varepsilon_{\mu\rho\alpha\beta}
p_1^\alpha p_2^\delta p_3^\gamma p_4^\beta,
\nonumber \\
{\mathcal{M}}^{(6b)}_{\mu\nu\lambda\rho} &=& \frac{g_{D^* D\rho}^{}
g_{\psi D^* D}^{}}{(p_1-p_3)^2 - M_D^2} \,
\varepsilon_{\nu\rho\gamma\delta} \varepsilon_{\mu\lambda\alpha\beta}
p_1^\alpha p_2^\gamma p_3^\beta p_4^\delta,
\nonumber \\
{\mathcal{M}}^{(6c)}_{\mu\nu\lambda\rho} &=& -\frac{g_{D^* D^* \rho}^{}
g_{\psi D^* D^*}^{}}{(p_2-p_3)^2 - M_{D^*}^2} \left\{ g^{\alpha\beta} -
\frac{(p_2-p_3)^\alpha (p_2-p_3)^\beta}{M_{D^*}^2} \right\}
\nonumber \\ && \mbox{} \times
\left[ 2 p_{3\nu} g_{\alpha\lambda} - (p_2+p_3)_\alpha g_{\nu\lambda} +
2
p_{2\lambda} g_{\nu\alpha} \right]
\left[ 2 p_{4\mu} g_{\beta\rho} - (p_1+p_4)_\beta g_{\mu\rho} + 2
p_{1\rho} g_{\mu\beta} \right],
\nonumber \\
{\mathcal{M}}^{(6d)}_{\mu\nu\lambda\rho} &=& -\frac{g_{D^* D^* \rho}^{}
g_{\psi D^* D^*}^{}}{(p_1-p_3)^2 - M_{D^*}^2} \left\{ g^{\alpha\beta} -
\frac{(p_1-p_3)^\alpha (p_1-p_3)^\beta}{M_{D^*}^2} \right\}
\nonumber \\ && \mbox{} \times
\left[ 2 p_{4\nu} g_{\alpha\rho} - (p_2+p_4)_\alpha g_{\nu\rho} + 2
p_{2\rho} g_{\nu\alpha} \right]
\left[ 2 p_{3\mu} g_{\beta\lambda} - (p_1+p_3)_\beta g_{\mu\lambda} + 2
p_{1\lambda} g_{\mu\beta} \right],
\nonumber \\
{\mathcal{M}}^{(6e)}_{\mu\nu\lambda\rho} &=& - g_{\psi D^* D^* \rho}^{}
\left( 2 g_{\mu\nu} g_{\lambda\rho} - g_{\mu\rho} g_{\nu\lambda} -
g_{\mu\lambda} g_{\nu\rho} \right).
\end{eqnarray}
In writing the above amplitudes, we have used $p_i \cdot \varepsilon_i = 0$.
Note that the cross section for $\pi(\rho) + J/\psi \to D + \bar{D}^*$ is
the same as that of  $\pi(\rho) + J/\psi \to D^* + \bar{D}$.

\section{Coupling Constants and Cross Sections}

\subsection{Coupling constants}

To estimate the cross sections, let us first determine the coupling
constants of our effective Lagrangian (\ref{Lag1}) and (\ref{Lag2}).
We follow the methods of Refs. \cite{Hagl00,LK00} to determine the
couplings for the normal interactions.
Here we briefly explain the method referring to Refs. \cite{Hagl00,LK00}
for details.
The coupling constant of $D^* D \pi$ can be determined from the
experimental data of $D^* \to D \pi$ decay.
Our effective Lagrangian gives
\begin{equation}
\Gamma_{D^* \to D \pi} = \frac{g_{D^* D \pi}^2}{24\pi}
\frac{|{\bf p}_\pi|^3}{M_{D^*}^2}.
\end{equation}
However, at present, only its upper bound ($\simeq 0.131$ MeV)
is known experimentally \cite{ACCMOR92,PDG00}, which translates into 
$g_{D^* D \pi}^{} \le 14.7$.
Therefore, we can only rely on model predictions such as those based on
the QCD sum rule approach \cite{BBKR95} or relativistic quark potential
model \cite{CDN94}.  
Here we will take the value obtained in Ref. \cite{BBKR95} and use%
\footnote{Note that our convention is different from that of Ref.
\cite{BBKR95} by a factor of $\sqrt{2}$.}
\begin{equation}
g_{D^* D \pi}^{} = 8.8.
\end{equation}

For $\psi DD$, $DD \rho$ and $\psi D^* D^*$, $D^* D^* \rho$ couplings,
we will follow Refs. \cite{MMu98,LK00} and make use of the vector meson
dominance model (VDM).
By applying VDM to the couplings of $\gamma D^+ D^-$ and $\gamma D^0
\bar{D}^0$, one obtains \cite{LK00}
\begin{equation}
g_{\psi DD}^{} = \frac23 g_{J/\psi}^{} = 7.71,
\qquad
g_{DD\rho}^{} = \frac12 g_{\rho}^{} = 2.52,
\end{equation}
where $g_V^{}$ is determined from
\begin{equation}
\Gamma_{V \to e^+e^-} = \frac{4\pi \alpha_{\rm em}^2}{3} \frac{M_V}{g_V^2},
\end{equation}
with the vector meson mass $M_V$ and $\alpha_{\rm em} = e^2/4\pi$.
The same method can be applied to the couplings of $\gamma D^{*+} D^{*-}$
and $\gamma D^{*0} \bar{D}^{*0}$.
This gives 
\begin{equation}
g_{\psi D^*D^*}^{} = g_{\psi DD}^{},
\qquad
g_{D^*D^*\rho}^{} = g_{DD\rho}^{}.
\end{equation}
It is shown in Ref. \cite{LK00} that the SU(4) symmetry values are not
far from the above estimates except $g_{\psi DD}^{}$.

There is no experimental or phenomenological informations on the 4-point
vertices. Hence, we will rely on the SU(4) symmetry relations.
Following Ref. \cite{LK00}, therefore, we assume that the appropriate
symmetry breaking effects in the 4-point coupling constants are encoded
via their relations to the 3-point vertices within SU(4) symmetry.   
Hence, using the phenomenological estimate of the 3-point vertices
fixed above, we have
\begin{eqnarray}
g_{\psi D^* D\pi}^{} &=& \frac{1}{2} g_{\psi DD}^{} g_{D^* D \pi}^{}
\approx 33.92,
\nonumber \\
g_{\psi DD \rho}^{} &=& 2 g_{\psi DD}^{} g_{DD\rho}^{} \approx 38.86,
\nonumber \\
g_{\psi D^* D^* \rho}^{} &=& g_{\psi D^*D^*}^{} g_{D^*D^* \rho}^{}
\approx 19.43.
\end{eqnarray}

We determine the couplings of the anomalous interactions in a similar
way.
As discussed before, the $D^* D^* \pi$ coupling is constrained by the
heavy quark spin symmetry.
Comparing with the effective Lagrangian of Ref. \cite{Chan97}, we get
to leading order in $1/\bar{M}(D)$ 
\begin{equation}
g_{D^* D^* \pi} = \frac{\bar{M}(D)}{2} g_{D^* D \pi}^{} \approx 9.08
\mbox{ GeV}^{-1},
\end{equation}
where $\bar{M}(D)$ is the average mass of $D$ and $D^*$.

For $\psi D^* D$ and $D^* D \rho$ couplings, we can apply VDM to the
radiative decays of $D^*$ into $D$, i.e., $D^* \to D \gamma$.
Then by using the same technique explained before, we obtain
\begin{eqnarray}
g_{\psi D^* D}^{} &=& \frac23 g_{J/\psi}^{} g_V^{\pm} = \frac23
g_{J/\psi}^{} g_V^{0}, \nonumber \\
g_{D^* D \rho}^{} &=& \frac16 g_\rho^{} \left( g_V^\pm + 2 g_V^0
\right),
\end{eqnarray}
where $g_V^{\pm,0}$ are defined through
\begin{eqnarray}
\langle D^+(k) | J_\mu^{em} | D^{*+}(p,\varepsilon) \rangle &=& e
g_V^\pm
\varepsilon_{\mu\nu\alpha\beta} \varepsilon^\nu k^\alpha p^\beta,
\nonumber \\
\langle \bar{D}^0(k) | J_\mu^{em} | \bar{D}^{*0}(p,\varepsilon) \rangle
&=& e g_V^0
\varepsilon_{\mu\nu\alpha\beta} \varepsilon^\nu k^\alpha p^\beta.
\end{eqnarray}
Then from the experimental values for the $D^*$ radiative decays, we can
determine the above coupling constants.
However, at present, only an experimental upper bound is known for the
$D^*$ decays.  
Therefore, we will use model prediction based on relativistic potential
model given in Ref. \cite{CDN94}, which gives
\begin{equation}
g_V^{} = \frac{e_Q^{}}{\Lambda_Q} + \frac{e_q^{}}{\Lambda_q},
\end{equation}
where $e_Q^{}$ is the heavy quark charge and $e_q^{}$ the light quark
charge with $\Lambda_Q = 1.57$ and $\Lambda_q = 0.48$ in GeV unit.
This gives
\begin{equation}
g_V^\pm \approx 1.12 \mbox{ GeV}^{-1}, \qquad
g_V^0 \approx 0.96 \mbox{ GeV}^{-1},
\end{equation}
which then leads to
\begin{equation}
g_{\psi D^* D}^{} = 7.40 \sim 8.64 \mbox{ GeV}^{-1}, \qquad
g_{D^* D \rho}^{} = 2.82 \mbox{ GeV}^{-1}.
\end{equation}

For the 4-point couplings, we depend on the SU(4) relations again,
which gives
\begin{eqnarray}
g_{\psi D^* D^* \pi}^{} &=& h_{\psi D^* D^* \pi}^{} =
\frac12 g_{\psi D^* D}^{} g_{D^* D \pi}
\approx 38.19 \mbox{ GeV}^{-1},
\nonumber \\
g_{\psi D^* D \rho}^{} &=& h_{\psi D^* D \rho}^{} =
g_{\psi D^* D}^{} g_{D^* D^* \rho}^{}
\approx 21.77 \mbox{ GeV}^{-1}.
\end{eqnarray}
However, for the $\psi DD \pi$ coupling, it is not easy to write it in
terms of the other 3-point coupling constants as can be seen from the
difference in dimensions.
Thus we directly use the SU(4) relation and assume that the symmetry
breaking effects change $F_\pi$ to $F_D$ \cite{PSWa91-OMRS91};
\begin{equation}
g_{\psi DD \pi}^{} = \frac{g_{D^* D \pi}^{}N_c}{6\pi^2F_D^3} \approx 16.0
\mbox{ GeV}^{-3}.
\end{equation}
Here we have used $F_D \approx 2.3 F_\pi$ \cite{PDG00}.
This completes the determination of our coupling constants.
The values of the coupling constants are listed in Table~\ref{tab:cc}
together with their fixing procedure.

\subsection{Cross sections without form factors}

\begin{figure}
\centering
\epsfig{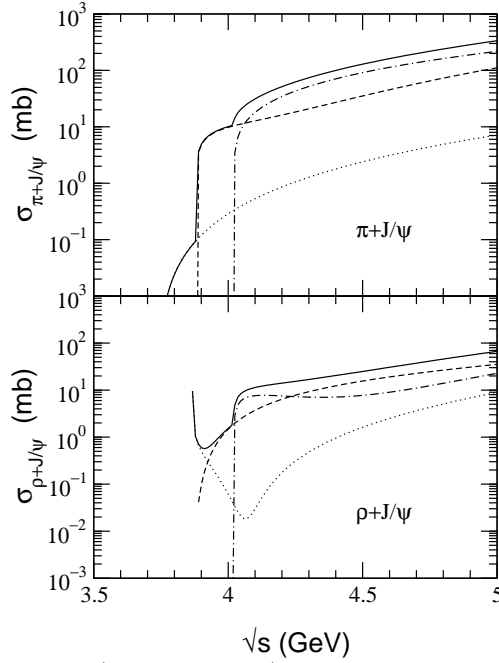}
\caption{Cross sections for $\pi + J/\psi$ and $\rho + J/\psi$.
In the upper panel, the dotted, dashed, and dot-dashed lines correspond to
Diagrams (1), (2), and (3), respectively. In the lower panel,
the dotted, dashed, and dot-dashed lines correspond to 
Diagrams (4), (5), and (6), respectively. The solid lines are the sum of
all processes.}
\label{fig:all}
\end{figure}

We first give the cross sections for the $J/\psi$ absorption processes
without any form factors.
Since our meson exchange model is not expected to work at high energies,
we focus on the energy region near threshold in this paper.
In Fig. \ref{fig:all} we show the cross sections for the $J/\psi$
absorption processes obtained from Figs. \ref{fig:pions} and \ref{fig:rhos}.
Although the $\pi + J/\psi \to D + \bar{D}$ process has the lowest
threshold energy, its cross section (dotted line in the upper panel of
Fig. \ref{fig:all}) is suppressed compared to the other channels over 
the whole energy region.  
Nevertheless, this process was not considered previously in the meson
exchange models.  
When the energy is less than the  $D^* + \bar{D}$ threshold, this process
alone is allowed, but the magnitude is very small and less than $0.1$ mb.
This observation is consistent with Ref. \cite{BBIK00}, although its energy
dependence is different.

The $\pi + J/\psi \to D + \bar{D}^*$ has the same cross section as the
$\pi + J/\psi \to D^* + \bar{D}$, and their sum is given by the dashed
line in the upper panel of Fig. \ref{fig:all}.
These processes dominate for the energies from their threshold up to
the $D^* + \bar{D}^*$ threshold.
At higher energies, the $\pi + J/\psi \to D^* + \bar{D}^*$ process
dominates the absorption.  
Since the $D^* + \bar{D}^*$ threshold is larger than $4$ GeV, the
$\pi + J/\psi \to D^* + \bar{D}, D + \bar{D}^*$ is the dominant process
for $\sqrt{s} \leq 4$ GeV.

The cross section for $\pi + J/\psi \to D^* + \bar{D}^*$ process (the
dot-dashed line in the upper panel of Fig. \ref{fig:all}) is the
dominant one at large energies, which is in contrast to the quark
exchange model calculations \cite{MBQ95,WSB00b}.
In addition to the effects from the form factors, this may indicate
that the meson exchange approach for this reaction has non-trivial
contributions from the exchanges of meson resonance with heavier mass,
such as the $D_1$ exchange, because of its higher threshold.
This process is allowed by the anomalous terms and was not considered
in the previous meson exchange model studies.

In the lower panel of Fig. \ref{fig:all} we show the cross sections
for the  $\rho + J/\psi$ processes.
It is shown that at very low energy, the exothermic $\rho + J/\psi
\to D + \bar{D}$ is the dominant process (the dotted line in the lower
panel of Fig. \ref{fig:all}).
But it is rapidly taken over by $\rho + J/\psi \to D^* + \bar{D}, D +
\bar{D}^*$ process (the dashed line in the lower panel of
Fig. \ref{fig:all}), which also dominates at higher energies.
This reaction is an anomalous parity process and was not studied before
in the meson exchange model.
The $\rho + J/\psi \to D^* + \bar{D}^*$ cross section (the dot-dashed line
in the lower panel of Fig.  \ref{fig:all}) has a peak near threshold
but has comparable size with that of $\rho + J/\psi \to D^* +
\bar{D}, D + \bar{D}^*$.

\begin{figure}
\centering
\epsfig{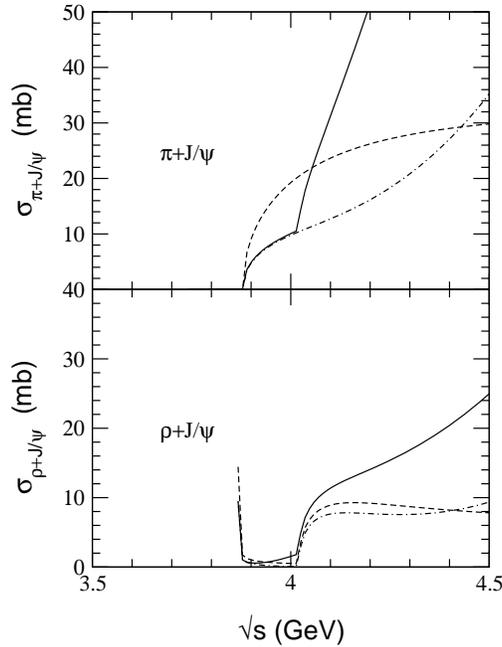}
\caption{Cross sections for $\pi + J/\psi$ and $\rho + J/\psi$.
The solid lines are our model predictions and the dashed lines are those
obtained without the anomalous terms. The dot-dashed lines correspond to
$\pi + J/\psi \to D^* + \bar{D}, D + \bar{D}^*$ in the upper panel and
$\rho + J/\psi \to D + \bar{D}, D^* + \bar{D}^*$ in the lower panel in
full calculation.}
\label{fig:compare}
\end{figure}

The dashed lines in Fig. \ref{fig:compare} show the results obtained with
the normal terms only and are consistent with the results of Refs.
\cite{Hagl00,LK00,HG00b}.
The solid lines are the predictions of the full calculation including the
anomalous terms.
It is evident that the anomalous terms give non-trivial effects to the
cross sections especially for the $\pi + J/\psi$ reaction.
For comparison, we also calculate the normal processes considered in previous
studies but with the additional diagrams with anomalous vertices.
The results are given by the dot-dashed lines in Fig. \ref{fig:compare}.  
The difference between the dashed lines and dot-dashed lines are due to
the anomalous processes, i.e., for $\pi + J/\psi \to D^* + \bar{D}, D +
\bar{D}^*$ in the upper panel and for $\rho + J/\psi \to D + \bar{D},
D^* + \bar{D}^*$ in the lower panel.  
At low energy, i.e., $\sqrt{s} \leq 4$ GeV which corresponds to
$E_{\rm kin} = \sqrt{s} - M_{J/\psi} - M_\pi \leq 0.8$ GeV, where
$M_{J/\psi}$ and $M_\pi$ are the $J/\psi$ and pion masses respectively,
the anomalous interactions reduce the total absorption cross section for
$\pi + J/\psi$ by about $50$~\%.
Our numerical results show that the $\pi + J/\psi$ absorption cross
section is less than $10$ mb in this energy region.
However, the contribution of the anomalous terms to the $\rho + J/\psi$
cross sections is not so important except at the high energy region,
which is thought to have many corrections from higher meson resonances
exchanges.

\begin{figure}
\centering
\epsfig{file=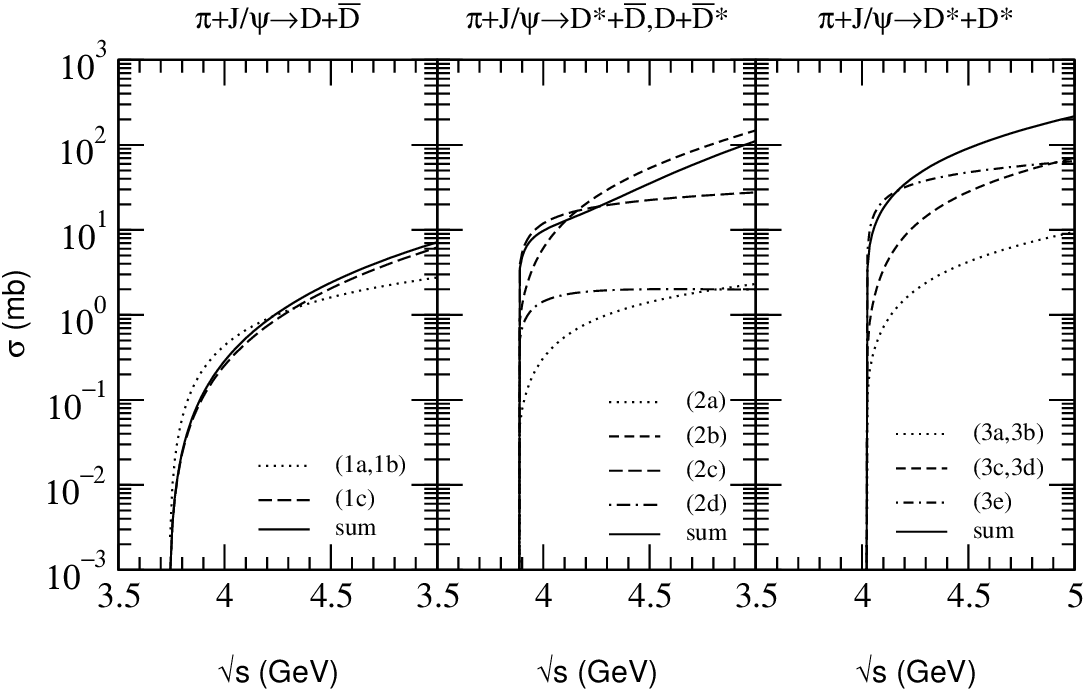, width=0.7\hsize}
\caption{Contributions from each diagram to the cross sections for
$\pi + J/\psi$.}
\label{fig:pionscs}
\end{figure}

\begin{figure}
\centering
\epsfig{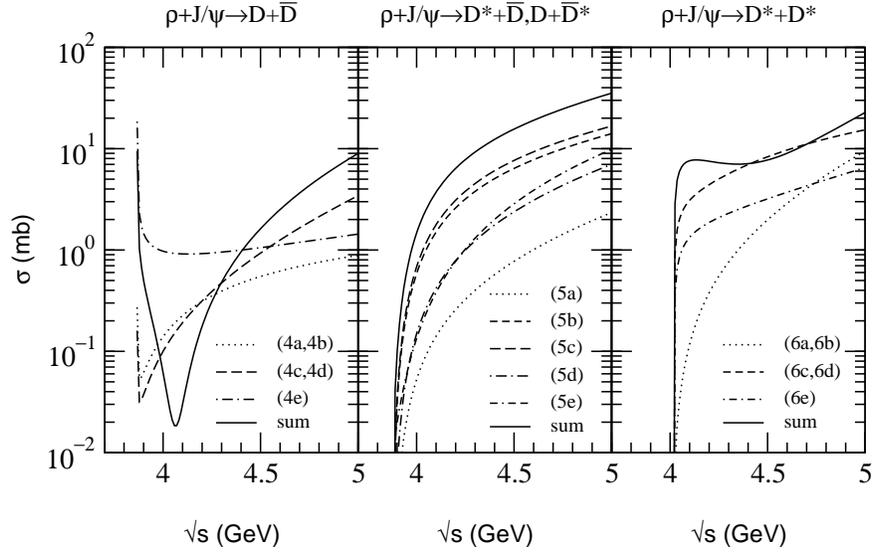}
\caption{Contributions from each diagram to the cross sections for
$\rho + J/\psi$.}
\label{fig:rhoscs}
\end{figure}

In Figs. \ref{fig:pionscs} and \ref{fig:rhoscs}, we show the contributions 
from each diagram for each channel.
For $\pi + J/\psi \to D + \bar{D}$ and $\pi + J/\psi \to D^* + \bar{D}^*$,
it can be seen that the contact diagrams (1c) and (3e) are important.
However, considering the uncertainty related to the 4-point vertices, 
the results for this process are more qualitative.
The more important and less uncertain process is the $\pi + J/\psi \to D^*
+ \bar{D}$.  
The interesting observation here is that the role of the anomalous term
[(2b) of Fig. \ref{fig:pions}] is quite non-trivial at low energies and
even dominates at higher energies.
The strong interference with the normal terms leads to the substantial
difference in the cross section as shown in the upper panel of Fig.
\ref{fig:compare}.    
But in contrast, as can be seen from Fig. \ref{fig:rhoscs}, the role of
the anomalous terms in the $\rho + J/\psi$ absorption is not so crucial
except $\rho + J/\psi \to D^* + \bar{D}, D + \bar{D}^*$ which is allowed
only by including the anomalous vertices.
Its effect becomes important at higher energies and is small at
$\sqrt{s} \leq 4$ GeV.

\subsection{Cross sections with form factors}

As pointed out by Ref. \cite{WSB00b}, the assumption of $t$-channel
exchange of a heavy meson is hard to justify without including form
factors.
This is so because the range of heavy meson exchange is much smaller
than the physical sizes of the initial hadrons.
There are some approaches to include form factors in the $J/\psi$
absorption processes and different form factors and cutoff parameters
are employed \cite{LK00,HG00b,HG00}, which, of course, cannot be
justified {\em a priori\/}.
Microscopic models such as quark potential models may give us a guide
for the form factors.
Here, we employ the simple form of the form factors employed in
Ref. \cite{LK00}, which are
\begin{equation}
F_3 = \frac{\Lambda^2}{\Lambda^2 + r^2}, \qquad
F_4 = \frac{\Lambda^2}{\Lambda^2 + \bar{r}^2} \frac{\Lambda^2}{\Lambda^2 +
\bar{r}^2},
\label{form34}
\end{equation}
where $r^2 = ({\bf p}_1-{\bf p}_3)^2$ or $({\bf p}_2-{\bf p}_3)^2$ and
$\bar{r} = [ ({\bf p}_1-{\bf p}_3)^2 + ({\bf p}_2-{\bf p}_3)^2]/2$.
$F_3$ is the form factor for the 3-point vertices and $F_4$ for the 4-point
vertices.
The cutoff parameters should be chosen experimentally and may take
different values for different vertices.
However, because of the paucity of experimental information, we use the
same value for all cutoff parameters as in the literature and investigate
the dependence of the cross sections on the cutoff parameters.

\begin{figure}
\centering
\epsfig{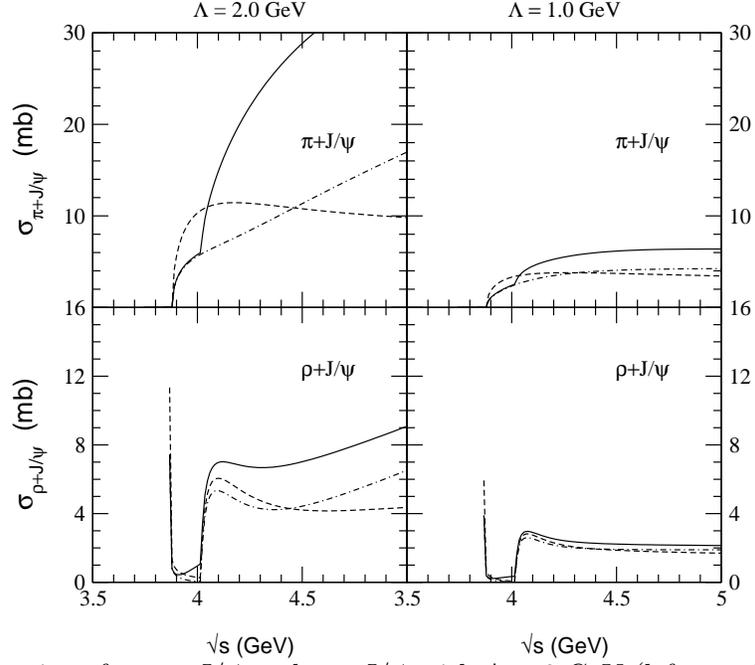}
\caption{Cross sections for $\pi + J/\psi$ and $\rho + J/\psi$ with
$\Lambda = 2$ GeV (left panel) and $1$ GeV (right panel).
The notations are the same as in Fig. \ref{fig:compare}.}
\label{fig:compall}
\end{figure}

In Fig. \ref{fig:compall} we show a similar figure as in
Fig. \ref{fig:compare} but with the form factors of Eq. (\ref{form34}) 
with $\Lambda = 2$ GeV (left panel) and $1$ GeV (right panel).
It can be seen that the role of the form factors are more important at
higher energy region and suppresses the overall cross sections.
But the tendency shown in Fig. \ref{fig:compare} is still valid, i.e.,
the anomalous terms reduce the cross sections for $\pi + J/\psi$ at low
energies through interference with the normal terms.
Finally, Fig. \ref{fig:compall} then shows that the cross sections for
$\pi + J/\psi$ absorption process is about $2 \sim 6$ mb depending on the
cutoff parameter at $\sqrt{s} \leq 4$ GeV.
The $\rho + J/\psi$ cross section is very small at $\sqrt{s} \leq 4$ GeV
and has maximum value of $3 \sim 9$ mb for intermediate energies up to
$5$ GeV.

\section{Summary and Discussions}

Knowing the magnitude and energy dependence of the $J/\psi$ absorption
cross sections by hadrons is crucial to estimate the effect of $J/\psi$
suppression in RHIC due to hadronic processes.      
We have re-examined the $J/\psi$ absorption by $\pi$ and $\rho$ mesons
in a meson exchange model.
The absorption processes are dominated by the $D$ and the $D^*$ exchanges.  
By including anomalous parity interactions we found that additional channels
and absorption mechanisms are allowed.
Their contribution was found to be quite non-trivial especially for
$\pi + J/\psi$ cross section.
The calculated cross section for $\pi + J/\psi$ at low energies was found
to be only about one half of the previous calculations
\cite{Hagl00,LK00,HG00b}.
With conventional, but arbitrarily chosen, form factors and cutoff
parameters, the $\pi + J/\psi$ cross section was estimated to be
$2 \sim 6$ mb for $\sqrt{s} \leq 4$ GeV.  
In contrast, the effect of anomalous terms was found to be weak in 
$\rho + J/\psi$ cross section at low energies.

In addition to the uncertainties related to the determination of the
coupling constants and form factors, our model calculation may be
improved further.  
One such improvement would be the inclusion of the axial-vector
$D_1(2420)$ meson exchanges, as mentioned in Ref. \cite{Hagl00}.
Also at low energy, final state interactions are expected to improve the
calculation.
It would be also interesting to investigate other possible absorption
channels as in Ref. \cite{HG00b}.
For example, one may consider OZI-evading vertices such as the
$J/\psi$-$\rho$-$\pi$ coupling, which can allow new absorption channels
with much smaller thresholds than the processes considered in this work,
although such processes are expected to be suppressed.


\acknowledgements

We are grateful to C.-Y. Wong for fruitful discussions and
encouragement.
This work was supported in part by the Brain Korea 21 project of Korean
Ministry of Education and by KOSEF under Grant No. 1999-2-111-005-5.


\appendix
\section*{}

In this Appendix we briefly explain the way to obtain the effective
Lagrangian (\ref{Lag1}) and (\ref{Lag2}).
As in the literature \cite{MMu98,Hagl00,LK00,HG00b} we start with the
SU(4) symmetric chiral Yang-Mills Lagrangian \cite{Mei88},
\begin{eqnarray}
{\mathcal{L}}_{\chi YM} &=&
\frac{F_\pi^2}{8} \,\mbox{Tr}\, \left( D_\mu U D^\mu U^\dagger \right)
+ \frac{F_\pi^2}{8} \,\mbox{Tr}\, \left[ M \left( U + U^\dagger - 2
\right) \right]
\nonumber \\ && \mbox{}
- \frac12 \,\mbox{Tr}\, \left( F_{\mu\nu}^L F^{L\mu\nu}
+ F_{\mu\nu}^R F^{R\mu\nu} \right)
+ m_0^2 \left( A_\mu^L A^{L\mu} + A_\mu^R A^{R\mu} \right),
\label{bareLag}
\end{eqnarray}
where $U = \exp(i2\phi/F_\pi)$ with the pion decay constant $F_\pi$
($\approx 132$ MeV) and $\phi = \pi^a \lambda^a / \sqrt{2}$.
The covariant derivative is defined as
\begin{equation}
D_\mu U = \partial_\mu U - i g A_\mu^L U + ig U A_\mu^R,
\end{equation}
with left- and right-handed spin-1 fields $A_\mu^{L,R}$, which are
related to the vector and axial-vector mesons $V_\mu$ ($=V_\mu^a
\lambda^a / \sqrt{2}$) and $A_\mu$ ($=A_\mu^a \lambda^a / \sqrt{2}$) by
\begin{equation}
A_{L\mu} = \frac12 \left( V_\mu + A_\mu \right), \qquad
A_{R\mu} = \frac12 \left( V_\mu - A_\mu \right).
\end{equation}
The field strength tensors are
\begin{equation}
F_{\mu\nu}^{L,R} = \partial_\mu A_\nu^{L,R} - \partial_\nu A_\mu^{L,R} -
ig [ A_\mu^{L,R} , A_\nu^{L,R} ].
\end{equation}
In obtaining the Lagrangian (\ref{bareLag}) we do not take any
non-minimal terms into account for simplicity.

It is straightforward to obtain the interaction Lagrangian for $\phi$
and $V$, which leads to
\begin{eqnarray}
{\mathcal{L}}^{VP}_{\rm n} &=&
- \frac{ig}{2} \,\mbox{Tr}\, \left\{ V_\mu \left( \phi
\partial^\mu \phi - \partial^\mu \phi \phi \right) \right\}
+ \frac{ig}{4} \,\mbox{Tr}\, ( \partial_\mu V_\nu - \partial_\nu V_\mu )
[ V^\mu, V^\nu ]
\nonumber \\ && \mbox{}
- \frac{g^2}{8} \,\mbox{Tr}\, [ V_\mu, \phi ] [V^\mu, \phi] 
+ \frac{g^2}{16} \,\mbox{Tr}\, [V_\mu, V_\nu]^2,
\label{bLag1}
\end{eqnarray}
where we have ignored the $A$-$\phi$ mixing effects as they are of order
of $1/M_V^2$.
The above Lagrangian was obtained by keeping the axial-vector field.
As in Ref. \cite{KS85}, one may gauge-out the axial-vector field.
It is found that including proper non-minimal terms gives the same
form for the effective Lagrangian as Eq. (\ref{bLag1}) with different
coupling constant \cite{HG00b}, i.e.,
by substituting $2g$ for the coupling constant of Ref. \cite{HG00b} one
can get the effective Lagrangian in the form of Eq. (\ref{bLag1}).

The anomalous parity terms come from nine gauged Wess-Zumino terms,
which read
\begin{eqnarray}
{\mathcal{L}}_{\rm an} &=& 5Ci \,\mbox{Tr}\, \left[ A_L L^3 + A_R R^3 \right]
- 5C \,\mbox{Tr}\, \left[ (dA_L A_L + A_L dA_L) L + (dA_R A_R +
A_R dA_R)R \right]
\nonumber \\ && \mbox{}
+ 5C \,\mbox{Tr}\, \left[ dA_L dU A_R U^\dagger - dA_R dU^\dagger
A_L U \right]
+ 5C \,\mbox{Tr}\, \left[ A_R U^\dagger A_L U R^2 - A_L U A_R
U^\dagger L^2 \right]
\nonumber \\ && \mbox{}
+ \frac{5C}{2} \,\mbox{Tr}\, \left[ (A_L L)^2 - (A_R R)^2 \right]
+ 5Ci \,\mbox{Tr}\, \left[ A_L^3 L + A_R^3 R \right]
\nonumber \\ && \mbox{}
+ 5Ci \,\mbox{Tr}\, \left[ \left( dA_R A_R + A_R dA_R \right)
U^\dagger A_L U - \left(dA_L A_L + A_L dA_L \right) U A_R U^\dagger
\right]
\nonumber \\ && \mbox{}
+ 5Ci \,\mbox{Tr}\, \left[ A_L U A_R U^\dagger A_L L + A_R
U^\dagger A_L U A_R R \right]
\nonumber \\ && \mbox{}
+ 5C \,\mbox{Tr}\, \left[ A_R^3 U^\dagger A_L U - A_L^3 U A_R U +
\frac12 (UA_R U^\dagger A_L)^2 \right],
\end{eqnarray}
where $C = -i N_c/240\pi^2$ with the number of color $N_c$ and $L = dU
U^\dagger$, $R = U^\dagger dU$.
We have used 1-form notation in writing the above Lagrangian.

For simplicity we again ignore the $A$-$\phi$ mixing and use the Bardeen
subtracted form for the anomalous terms.
This gives
\begin{eqnarray}
{\mathcal{L}}^{VP}_{\rm an} &=&
- \frac{g^2 N_c}{16 \pi^2 F_\pi} \,\mbox{Tr}\, \left( dV dV
\phi \right)
- \frac{igN_c}{6\pi^2 F_\pi^3} \,\mbox{Tr}\, \left\{ V
(d\phi)^3 \right\}
\nonumber \\ && \mbox{}
+ \frac{ig^3 N_c}{32 \pi^2 F_\pi} \,\mbox{Tr}\, \left( V^3
d\phi \right)
+ \frac{ig^3 N_c}{32\pi^2 F_\pi} \,\mbox{Tr}\, \left( V dV V \phi
\right).
\end{eqnarray}

Writing the pseudoscalar and vector fields explicitly and assuming pure
$\bar{c}c$ for the $J/\psi$ wave function, we can obtain the effective
Lagrangian (\ref{Lag1}) and (\ref{Lag2}) as well as the SU(4) symmetry
relations used in Sec.~III.
Note that by eliminating the axial-vector fields, the coupling constant
of $V\phi\phi\phi$ interaction becomes \cite{KS85}
\begin{equation}
g_{V\phi\phi\phi}^{} = \frac{M_V^2}{\pi^2 g_{V\phi\phi}^{} F_\pi^5}.
\end{equation}
Assuming that the proper symmetry breaking terms make $M_V \to M_{D^*}$,
$g_{V\phi\phi}^{} \to g_{D^*D \pi}^{}$, and $F_\pi \to F_D$ in the above
relation, we get $g_{\psi DD \pi}^{} \approx 18.0$ GeV$^{-3}$, which is
close to the value listed in Table \ref{tab:cc}.



\begin{table}[t]
\centering
\begin{tabular}{ccc} 
coupling constant & method & value \\ \hline
$g_{D^* D \pi}^{}$ & $D^* \to D \pi$ and QCD sum rule & $8.84$
\cite{BBKR95} \\
$g_{\psi DD}^{}$ & $\gamma DD$ coupling and VDM & $7.71$ \cite{MMu98} \\
$g_{DD \rho}^{}$ & $\gamma DD$ coupling and VDM & $2.52$ \cite{MMu98} \\
$g_{\psi D^*D^*}^{}$ & $\gamma D^* D^*$ coupling and VDM & $7.71$
\cite{MMu98,LK00} \\
$g_{D^*D^* \rho}^{}$ & $\gamma D^* D^*$ coupling and VDM & $2.52$
\cite{MMu98,LK00} \\
$g_{\psi D^* D \pi}^{}$ & SU(4) symmetry relation & $33.92$ \cite{LK00} \\
$g_{\psi D D \rho}^{}$ & SU(4) symmetry relation & $38.86$ \cite{LK00} \\
$g_{\psi D^* D^* \rho}^{}$ & SU(4) symmetry relation & $19.43$ \cite{LK00} \\
\hline
$g_{D^* D^* \pi}^{}$ & heavy quark spin symmetry & $9.08$ GeV$^{-1}$ \\
$g_{\psi D^* D}^{}$ & $D^* \to D \gamma$, VDM, and quark model & $8.64$
GeV$^{-1}$ \\
$g_{D^* D \rho}^{}$ & $D^* \to D \gamma$, VDM, and quark model & $2.82$
GeV$^{-1}$ \\
$g_{\psi DD \pi}^{}$ & SU(4) symmetry relation & $16.00$ GeV$^{-3}$  \\
$g_{\psi D^* D^* \pi}^{}$ & SU(4) symmetry relation &  $38.19$ GeV$^{-1}$ \\
$h_{\psi D^* D^* \pi}^{}$ & SU(4) symmetry relation &  $38.19$ GeV$^{-1}$ \\
$g_{\psi D^* D \rho}^{}$ & SU(4) symmetry relation &  $21.77$ GeV$^{-1}$ \\
$h_{\psi D^* D \rho}^{}$ & SU(4) symmetry relation &  $21.77$ GeV$^{-1}$ \\
\end{tabular}
\bigskip
\caption{Determination of the coupling constants.}
\label{tab:cc}
\end{table}


\begin{references}

\bibitem{MS86}
T.~Matsui and H.~Satz,
   Phys. Lett. B {\bf 178}, 416 (1986).

\bibitem{Satz00}
H.~Satz,
   Rep. Prog. Phys. {\bf 63}, 1511 (2000);
   Univ. Bielefeld Report No. BI-TP-2000/31, 
   hep-ph/0009099.

\bibitem{NA50-96}
\mbox{NA50 Collaboration,} M.~Gonin {\em et al.\/},
   Nucl. Phys. {\bf A610}, 404c (1996).

\bibitem{NA50-00}
\mbox{NA50 Collaboration,} M.~C. Abreu {\em et al.\/},
   Phys. Lett. B {\bf 477}, 28 (2000).

\bibitem{BO96a}
J.-P. Blaizot and J.-Y. Ollitrault,
   Phys. Rev. Lett. {\bf 77}, 1703 (1996).

\bibitem{Wong98a}
C.-Y. Wong,
   Nucl. Phys. {\bf A630}, 487c (1998);
   Talk at 3rd Catania Relativistic Ion Studies on Phase Transitions
   Strong Interactions: Status and Perspectives (CRIS 2000),
   Italy, May, 2000, nucl-th/0007046.

\bibitem{comove}
J.~Ft\'{a}\v{c}nik, P.~Lichard, and J.~Pi\v{s}\'{u}t,
   Phys. Lett. B {\bf 207}, 194 (1988); 
S.~Gavin, M.~Gyulassy, and A.~Jackson, 
   {\em ibid.\/} {\bf 207}, 257 (1988); 
R.~Vogt, M.~Prakash, P.~Koch, and T.~H. Hansson, 
   {\em ibid.\/} {\bf 207}, 263 (1988).

\bibitem{CKo97}
W.~Cassing and C.~M. Ko,
   Phys. Lett. B {\bf 396}, 39 (1997).

\bibitem{CB97}
W.~Cassing and E.~L. Bratkovskaya,
   Nucl. Phys. {\bf A623}, 570 (1997).

\bibitem{AC98}
N.~Armesto and A.~Capella,
   Phys. Lett. B {\bf 430}, 23 (1998);
A. Capella, E. G. Ferreiro, and A. B. Kaidalov,
   Phys. Rev. Lett. {\bf 85}, 2080 (2000);
A. Sibirtsev, K. Tsushima, K. Saito, and A. W. Thomas,
   Phys. Lett. B {\bf 484}, 23 (2000).

\bibitem{KSa94}
D.~Kharzeev and H.~Satz,
   Phys. Lett. B {\bf 334}, 155 (1994).

\bibitem{KSSZ96}
D.~Kharzeev, H.~Satz, A.~Syamtomov, and G.~Zinovjev,
   Phys. Lett. B {\bf 389}, 595 (1996).

\bibitem{MBQ95}
K.~Martins, D.~Blaschke, and E.~Quack,
   Phys. Rev. C {\bf 51}, 2723 (1995).

\bibitem{WSB00b}
C.-Y. Wong, E.~S. Swanson, and T.~Barnes,
   Phys. Rev. C {\bf 62}, 045201 (2000).

\bibitem{MMu98}
S.~G. Matinyan and B.~M{\"u}ller,
   Phys. Rev. C {\bf 58}, 2994 (1998).

\bibitem{Hagl00}
K.~L. Haglin,
   Phys. Rev. C {\bf 61}, 031902 (2000).

\bibitem{LK00}
Z.~Lin and C.~M. Ko,
   Phys. Rev. C {\bf 62}, 034903 (2000).

\bibitem{HG00b}
K.~L. Haglin and C.~Gale,
   St. Cloud State Univ. Report (2000), nucl-th/0010017.

\bibitem{Peskin79}
M.~E. Peskin, 
   Nucl. Phys. {\bf B156}, 365 (1979);
G. Bhanot and  M.~E. Peskin, 
   {\em ibid.\/} {\bf B156}, 391 (1979). 

\bibitem{Wise92}
M.~B. Wise,
   Phys. Rev. D {\bf 45}, 2188 (1992).

\bibitem{YCCL92}
T.-M. Yan, H.-Y. Cheng, C.-Y. Cheung, G.-L. Lin, Y.~C. Lin, and H.-L. Yu,
   Phys. Rev. D {\bf 46}, 1148 (1992), {\bf 55}, 5851(E) (1997).

\bibitem{VS87-88}
M.~B. Voloshin and M.~A. Shifman,
  Yad. Fiz. {\bf 45}, 463 (1987),
  [Sov. J. Nucl. Phys. {\bf 45}, 292 (1987)];
  {\em ibid.\/} {\bf 47}, 801 (1988),
  [Sov. J. Nucl. Phys. {\bf 47}, 511 (1988)].

\bibitem{IW89-90}
N.~Isgur and M.~B. Wise,
   Phys. Lett. B {\bf 232}, 113 (1989);
   {\bf 237}, 527 (1990).

\bibitem{Chan97}
L.-H. Chan,
   Phys. Rev. D {\bf 55}, 5362 (1997).

\bibitem{VRIs}
M.~Luke and A.~V. Manohar,
   Phys. Lett. B {\bf 286}, 348 (1992);
Y.-Q. Chen,
   {\em ibid.\/} {\bf 317}, 421 (1993);
M.~Finkemeier, H.~Georgi, and M.~McIrvin,
   Phys. Rev. D {\bf 55}, 6933 (1997).

\bibitem{MOPR95}
D.-P. Min, Y.~Oh, B.-Y. Park, and M.~Rho,
   Int. J. Mod. Phys. E {\bf 4}, 47 (1995).

\bibitem{Mei88}
See, e.g., U.-G. Meissner,
   Phys. Rep. {\bf 161}, 213 (1988).

\bibitem{KS85}
{\"{O}}.~Kaymakcalan and J.~Schechter,
   Phys. Rev. D {\bf 31}, 1109 (1985).

\bibitem{KRS84}
{\"{O}}.~Kaymakcalan, S.~Rajeev, and J.~Schechter,
   Phys. Rev. D {\bf 30}, 594 (1984).

\bibitem{BBIK00}
D.~B. Blaschke, G.~R.~G. Burau, M.~A. Ivanov,
\mbox{Yu.} L.~Kalinovsky, and P.~C. Tandy,
   Univ. Rostock Report No. MPG-VT-UR-201-00,
   hep-ph/0002047.

\bibitem{ACCMOR92}
\mbox{ACCMOR Collaboration,} S.~Barlag {\em et al.\/},
   Phys. Lett. B {\bf 278}, 480 (1992).

\bibitem{PDG00}
\mbox{Particle Data Group,} D.~E.~Groom {\em et~al.\/},
   Eur. Phys. J. C {\bf 15}, 1 (2000).

\bibitem{BBKR95}
V.~M. Belyaev, V.~M. Braun, A.~Khodjamirian, and R.~R{\"u}ckl,
   Phys. Rev. D {\bf 51}, 6177 (1995).

\bibitem{CDN94}
P.~Colangelo, F.~De~Fazio, and G.~Nardulli,
   Phys. Lett. B {\bf 334}, 175 (1994).

\bibitem{PSWa91-OMRS91}
G.~Pari, B.~Schwesinger, and H.~Walliser,
   Phys. Lett. B {\bf 255}, 1 (1991);
Y.~Oh, D.-P. Min, M.~Rho, and N.~N. Scoccola,
   Nucl. Phys. {\bf A534}, 493 (1991).

\bibitem{HG00}
K.~Haglin and C.~Gale,
   in {\em Hirschegg 2000: Hadrons in Dense Matter\/},
   edited by M.~Buballa, W.~N{\"o}renberg, B.-J. Schaefer,
   and J.~Wambach, (GSI, Darmstadt, 2000),
   nucl-th/0002029.

\end{references}
\end{document}